\documentclass[a4paper,11pt]{article}
\usepackage{pos}

\usepackage{listings}

\newcommand{\GMTCalc}{\texttt{GM2Calc}}

\newcommand{\GAMBIT}{\texttt{GAMBIT}}
\newcommand{\mathematica}{\texttt{Ma\-the\-ma\-ti\-ca}}
\newcommand{\python}{\texttt{Py\-thon}}

\newcommand{\amu}{\ensuremath{a_\mu}}
\newcommand{\amuB}{\ensuremath{\amu^{\text{B}}}}
\newcommand{\amuF}{\ensuremath{\amu^{\text{F}}}}
\newcommand{\amuBSM}{\ensuremath{\amu^{\BSM}}}

\newcommand{\tl}{\ensuremath{{2\ell}}}

\newcommand{\hc}{\text{h.\,c.}}

\newcommand{\PiL}{\Pi^0} 

\newcommand{\VCKM}{V_{\scalefont{.9}\text{CKM}}}
\newcommand{\SM}{{\scalefont{.9}\text{SM}}}
\newcommand{\BSM}{{\scalefont{.9}\text{BSM}}}
\newcommand{\MSSM}{{\scalefont{.9}\text{MSSM}}}
\newcommand{\THDM}{{\scalefont{.9}\text{2HDM}}}
\newcommand{\FATHDM}{{\scalefont{.9}\text{FA2HDM}}}

\newcommand{\unit}[1]{\,\text{#1}}
\newcommand{\GeV}{\unit{GeV}}

\renewcommand{\imath}{\text{i}}

\title{GM2Calc - 2 for the 2HDM}

\author[a,b]{Peter Athron}
\author[b]{Csaba Balazs}
\author[c]{Adriano Cherchiglia}
\author*[b]{Douglas Jacob}
\author[d]{Dominik Stöckinger}
\author[d]{Hyejung Stöckinger-Kim}
\author[e]{Alexander Voigt}
\affiliation[a]{Department  of  Physics  and  Institute  of  Theoretical  Physics,  Nanjing  Normal  University, Nanjing, Jiangsu 210023, China}
\affiliation[b]{ARC Centre of Excellence for Particle Physics at
	the Terascale, School of Physics, Monash University, Melbourne,
	Victoria 3800, Australia}
\affiliation[c]{Centro de Ci\^ecias Naturais e Humanas, Universidade Federal do ABC, Santo Andr\'e, Brazil}
\affiliation[d]{Institut für Kern- und Teilchenphysik,
	TU Dresden, Zellescher Weg 19, 01069 Dresden, Germany}
\affiliation[e]{Fachbereich Energie und Biotechnologie, Hochschule Flensburg, Kanzleistraße 91--93, 24943 Flensburg, Germany}

\emailAdd{douglas.jacob@monash.edu}

\abstract{\GMTCalc\ is a leading tool for calculating precise contributions to $\amu$ in the Minimal Supersymmetric Standard Model.  In this proceeding we detail \GMTCalc\ version 2 where it is extended so it can calculate two-loop contributions to $\amu$ in the Two-Higgs Doublet Model (\THDM), based on the work in Ref.\ \cite{Athron:2021evk}.  The \THDM\ is a simple model, yet it is one of the few single field extensions of the Standard Model which is able to explain the muon $g-2$ anomaly.  We demonstrate the powerful and flexible \THDM\ capabilities of \GMTCalc2, which include the most precise contributions in the literature and allow the user to work in their favourite type of the \THDM\ as well as use complex and lepton flavour violating couplings.  With its multiple interfaces and input flexibility, \GMTCalc2 is a powerful tool both as a standalone code and as part of a larger code toolchain.  }

\FullConference{%
  Computational Tools for High Energy Physics and Cosmology (CompTools2021)\\
  22-26 November 2021\\
  Institut de Physique des 2 Infinis (IP2I), Lyon, France
}

\begin{document}
\maketitle

\section{Introduction} \label{sec:intro}

Currently there exists a $4.2\sigma$ discrepancy between the Standard Model (SM) and experimental values of $\amu$.  This discrepancy was confirmed by the Fermilab experiment \cite{PhysRevLett.126.141801}, who measured $\amu$ to be
\begin{equation} \label{eqn:Gm2Exp}
	\amu^\text{Exp} = (11659206.1\pm4.1)\times10^{-10},
\end{equation}
after combining their results with those from Brookhaven National Laboratory \cite{Tanabashi2018}.  This is higher than the SM prediction from the Muon $g-2$ Theory Initiative White Paper \cite{aoyama2020anomalous} (which itself builds on results done in Refs.\  \cite{Czarnecki:2002nt,Gnendiger:2013pva,Davier:2017zfy,Keshavarzi:2018mgv,Colangelo:2018mtw,Hoferichter:2019gzf,Davier:2019can,Keshavarzi:2019abf,Kurz:2014wya,Melnikov:2003xd,Masjuan:2017tvw,Colangelo:2017fiz,Hoferichter:2018kwz,Gerardin:2019vio,Bijnens:2019ghy,Colangelo:2019uex,Pauk:2014rta,Danilkin:2016hnh,Jegerlehner:2017gek,Knecht:2018sci,Eichmann:2019bqf,Roig:2019reh,Blum:2019ugy,Colangelo:2014qya})
\begin{equation} \label{eqn:Gm2SM}
	\amu^{\SM} = (11659181.0\pm4.3)\times10^{-10},
\end{equation}
by the amount
\begin{equation} \label{eq:amuBSM}
	\amuBSM = (25.1\pm5.9)\times10^{-10}.
\end{equation} 
The size of this discrepancy, which is on order of the weak contributions to $\amu$ \cite{Czarnecki2001}, suggests new physics at work in the value of $\amuBSM$.  Many models have been proposed as an explanation of the $\amu$ discrepancy (for a review see \cite{Athron:2021iuf}), including the Two-Higgs Doublet Model (\THDM).  

The \THDM\ involves the simple addition of a 2nd Higgs doublet to the SM with identical properties to the first.  For a simple extension, it has been shown to be adept at explaining many different physics anomalies.  Type-II \THDM\ is motivated by the Minimal Supersymmetric Standard Model (\MSSM) (see e.g. Ref.\ \cite{Martin1997}).  The general or type-III \THDM\ has been shown to be able to explain the flavour changing charged current for $R(D)$, whilst respecting constraints from the lepton flavour violating (LFV) decays $\tau\to\mu\gamma$ and $\tau\to3\mu$ \cite{Athron:2021auq}. The \THDM\ may also provide an explanation for things such as LHC data \cite{Haber:2015pua,Baglio:2014nea,Eberhardt:2013uba,Chowdhury:2015yja,Chowdhury:2017aav}, B-anomalies \cite{Misiak:2017bgg,Hu:2016gpe,Li:2014fea,Enomoto:2015wbn,Arnan:2017lxi}, and decays of Higgs bosons \cite{Krause:2016oke,Krause:2016xku,Altenkamp:2017kxk,Altenkamp:2017ldc,Altenkamp:2018hrq,Denner:2018opp,Kanemura:2018yai,Krause:2019qwe,Aiko:2021can}, to name a few.  

There have several demonstrations of the ability of the \THDM\ to explain the value of $\amuBSM$, and a detailed list has been included in the original work in Ref.\ \cite{Athron:2021evk}.  Both type-X and flavour-aligned \THDM\ are shown to be able to explain the value of $\amuBSM$ in the review in Ref.\ \cite{Athron:2021iuf}.  Ref.\ \cite{Jueid:2021avn} studied the type-X \THDM\ and found that while it was possible to produce an explanation of $\amuBSM$ that was consistent with the electron anomalous magnetic moment and collider searches, it was not consistent with $\tau$ and $Z$ decays.  Ref.\ \cite{Han:2021gfu} also investigated the simultaneous explanation of the muon and electron magnetic moment anomalies in the type-X \THDM\, and found it was possible with a light CP-even scalar.  Ref.\ \cite{Dey:2021pyn} also found that the type-X \THDM\ can explain the $\amu$ anomaly whilst allowing for highscale completion in scenarios with a low-mass CP-odd scalar.  Refs.\ \cite{Hou:2021qmf,Hou:2021wjj} use LFV one-loop contributions in the General \THDM\ to explain the value of $\amuBSM$, while Ref.\ \cite{Athron:2021auq} showed the inconsistency of a simultaneous explanation of the $\amu$ anomaly with Flavour-Changing Neutral Currents in the General \THDM.  

Previously, \GMTCalc\ was a tool solely for calculating the contributions to $\amu^{\textrm{SUSY}}$ in the \MSSM.  It has been used to analyze the \MSSM\ in a variety of different scenarios (see e.g. \cite{Han:2016gvr,Ren:2017ymm,Endo:2020mqz,Chakraborti:2020vjp,Chakraborti:2021kkr,Chakraborti:2021dli,Wang:2021bcx,Endo:2021zal,Ibe:2021cvf,VanBeekveld:2021tgn,Aboubrahim:2021xfi,Chakraborti:2021bmv,Jeong:2021qey,Lamborn:2021snt}).  It was also included in \GAMBIT\ \cite{Cornell:2016gho} as a backend, where it was used in several global fits of the \MSSM\ shown in Refs.\ \cite{GAMBIT:2017zdo,GAMBIT:2017snp,Abdughani:2019wuv,Athron:2021iuf}.  We recently extended it to enable it to calculate contributions in the \THDM\ \cite{Athron:2021evk}.  We incorporated state-of-the-art calculations in \GMTCalc, allowing it to calculate contributions and return the uncertainty at up to the two-loop level.  In this proceeding we provide a quick overview of the \GMTCalc2's new \THDM\ capabilities.  

\section{Usage} \label{sec:usage}

\GMTCalc\ version 2 features $\amu$ contributions at the one-loop level taken from \cite{Athron:2021auq} and two-loop contributions taken from \cite{Cherchiglia:2016eui}.  See \cite{Athron:2021evk} for in-depth details on the contributions included.  Here we give an essential overview.

When extending \GMTCalc\ to include the \THDM\, we aimed to make the internal calculations as general as possible, so that it is possible to use the software with any of the popular types of the \THDM.
The general potential of the two Higgs doublets $\Phi_{1,2}$ in the lambda basis is given by
\begin{align} \label{eqn:lambdapotential}
	\begin{split}
		-\mathcal{L}_\text{Scalar} ={}&
		m_{11}^2 \Phi_1^\dagger \Phi_1
		+ m_{22}^2 \Phi_2^\dagger \Phi_2
		- \left[ m_{12}^2 \Phi_1^\dagger \Phi_2 + \hc \right] \\
		&+ \frac{1}{2} \lambda_1 \left( \Phi_1^\dagger \Phi_1 \right)^2
		+ \frac{1}{2} \lambda_2 \left( \Phi_2^\dagger \Phi_2 \right)^2
		+ \lambda_3 \left( \Phi_1^\dagger \Phi_1 \right)\left( \Phi_2^\dagger \Phi_2 \right)
		+ \lambda_4 \left( \Phi_1^\dagger \Phi_2 \right)\left( \Phi_2^\dagger \Phi_1 \right) \\
		&+ \left[
		\frac{1}{2} \lambda_5 \left( \Phi_1^\dagger \Phi_2 \right)^2
		+ \lambda_6 \left( \Phi_1^\dagger \Phi_1 \right)\left( \Phi_1^\dagger \Phi_2 \right)
		+ \lambda_7 \left( \Phi_2^\dagger \Phi_2 \right)\left( \Phi_1^\dagger \Phi_2 \right)
		+ \hc
		\right].  
	\end{split} 
\end{align}
In addition to $\tan{(\beta)}$, we can use the lambdas $\lambda_{1,\ldots,5}$ and $m_{12}^2$ as input parameters for our \GMTCalc\ calculation, or we can transform the components of the two Higgs doublets into their mass eigenstates, and use the masses $m_{h,H,A,H^\pm}$, $m_{12}^2$, and the alignment parameter $\sin{(\beta-\alpha)}$ instead.  The quartic couplings $\lambda_{6,7}$ are not often non-zero in the literature, however \GMTCalc\ allows them to be for generality.  

In general the two Higgs doublets may have any of the below couplings to the \SM\ fermions
\begin{align} \label{eqn:yukawa}
	-\mathcal{L}_\text{Yuk} ={}&
	\Gamma_d ^0\overline{q_L ^0}\Phi_1 d_R ^0 + \Gamma_u ^0 \overline{q_L ^0}\Phi_1 ^{c} u_R ^0 + \Gamma_l ^0 \overline{l_L ^0}\Phi_1 e_R ^0
	+ \PiL_d \overline{q_L ^0}\Phi_2 d_R ^0 + \PiL_u \overline{q_L ^0}\Phi_2 ^{c} u_R ^0 + \PiL_l \overline{l_L ^0}\Phi_2 e_R ^0 + \hc,
\end{align}
where $\Phi_{1,2}^{c}=i\sigma_2\Phi_{1,2}^{*}$ The above $3\times3$ Yukawa matrices are allowed to take any complex value in the general \THDM.  We are able obtain the other types by setting the Yukawa matrices to the values shown in Table \ref{tab:types}.  If the user should choose one of the types-I, II, X or Y, then the Yukawa matrices are set according to these matrices (and the remaining ones according to the \SM\ masses), so then the user does not need to specify any Yukawa couplings.  In the flavour-aligned \THDM\ (\FATHDM) the ratio of the Yukawa matrices $\Pi_f^0$ and $\Gamma_f^0$ are fixed to be $\xi_f^{(*\textrm{ if }f=u)}$.  The user may input the strength of the Yukawa couplings in this case through the variable $\zeta_f$, which is related to the alignment parameter by
\begin{equation} \label{eqn:alignnment}
	\xi_f = \frac{\zeta_f + \tan{\beta}}{1 - \zeta_f \tan{\beta}}
\end{equation}
In the traditional \FATHDM\ there is no flavour violation (outside of the CKM matrix $\VCKM = V_u V_d^\dagger$, where $M_f^D = V_f M_f U_f^\dagger$ diagonalize the mass matrices), however, in \GMTCalc\ we allow the user to input a matrix of deviations from the diagonal $\Delta_f$.  Finally for the general \THDM\ we allow the user to directly input the matrices $\Pi_f = V_f \Pi_f^0 U_f^\dagger$.  This completes the list of parameters we need to calculate the contributions to $\amuBSM$ in \GMTCalc.  

\begin{table}[tb]
	\centering
	\caption{Values of the Yukawa matrices to recover the types-I, II, X, Y, and \FATHDM.}
	\begin{tabular}{l|c}
		\hline
		Type I & $\Gamma_u^0 = \Gamma_d^0 = \Gamma_l^0 = 0$ \\
		Type II & $\Gamma_u^0 = \Pi_d^0 = \Pi_l^0 = 0$ \\
		Type X & $\Gamma_u^0 = \Gamma_d^0 = \Pi_l^0 = 0$ \\
		Type Y & $\Gamma_u^0 = \Pi_d^0 = \Gamma_l^0 = 0$ \\
		Flavour-Aligned & $\Pi_u^0 = \xi_u^{*} \Gamma_u^0, \quad \Pi_{d,l}^0 = \xi_{d,l} \Gamma_{d,l}^0$ \\
		\hline
	\end{tabular}
	\label{tab:types}
\end{table}

\begin{figure}[tb]
	\centering
	\includegraphics[width=0.32\textwidth]{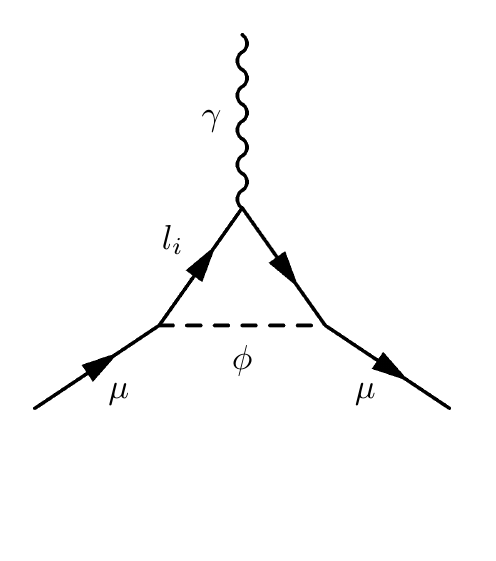}
	\includegraphics[width=0.32\textwidth]{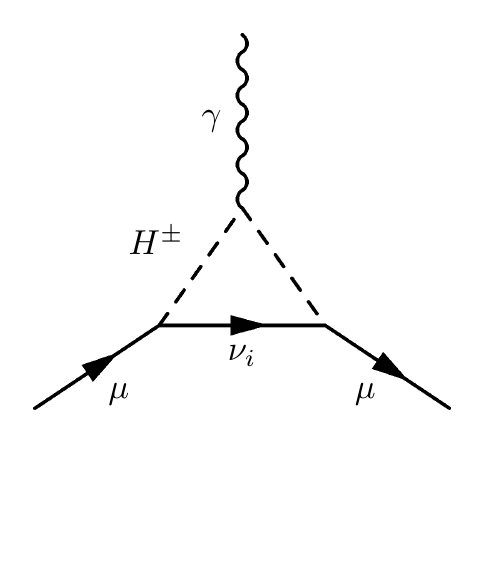}
	\caption{One-loop contribution included within \GMTCalc, where $\phi = h,H,A$ and $i=1,2,3$.  
	\label{fig:OneLoopDiagrams}}
\end{figure}

We divide the contributions into two parts, one-loop and two-loop
\begin{equation} \label{eqn:amufull}
	\amuBSM = \amu^{1\ell} + \amu^{2\ell}.  
\end{equation}
The one-loop contributions $\amu^{1\ell}$ included in $\GMTCalc$ are the full \BSM\ contributions and are shown in Fig.\ \ref{fig:OneLoopDiagrams}, and are defined in Eq.\ (40) of \cite{Athron:2021evk}.  They include diagrams with the \SM-like Higgs, since in the \THDM\ it can have non-\SM\ effects such as lepton flavour violation (which can occur if we include off-diagonal couplings in the matrix $\Delta_l$ in the \FATHDM\ or $\Pi_l$ in the general \THDM) and Higgs alignment.  To make sure that we are only including \BSM\ contributions we subtract off the \SM-like Higgs contributions that would be included in the \SM\ prediction.  

\begin{figure}[tb]
	\centering
	\includegraphics[width=0.32\textwidth]{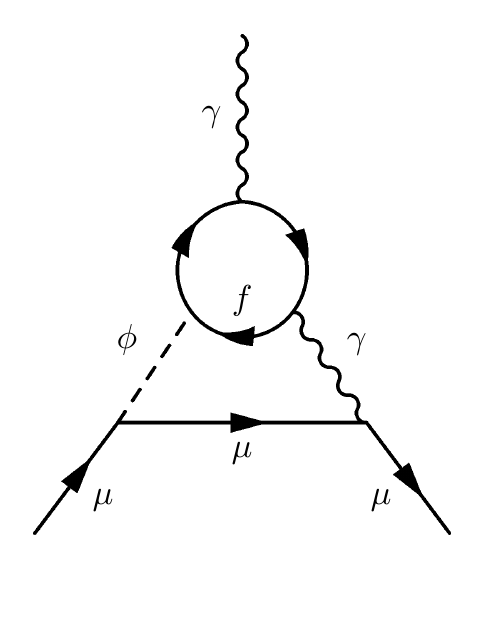}\hfill
	\includegraphics[width=0.32\textwidth]{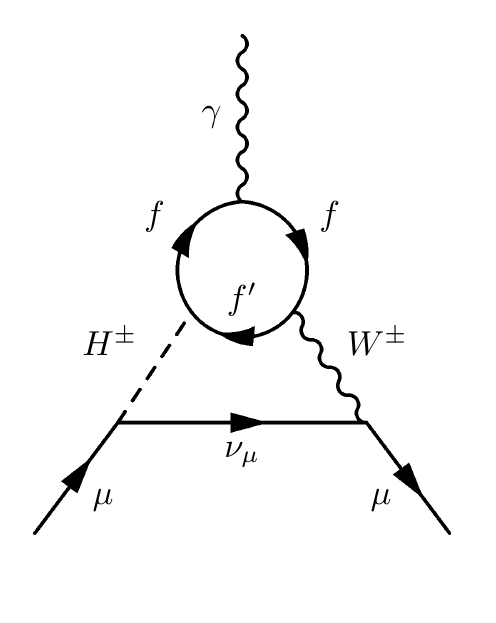}\hfill
	\includegraphics[width=0.32\textwidth]{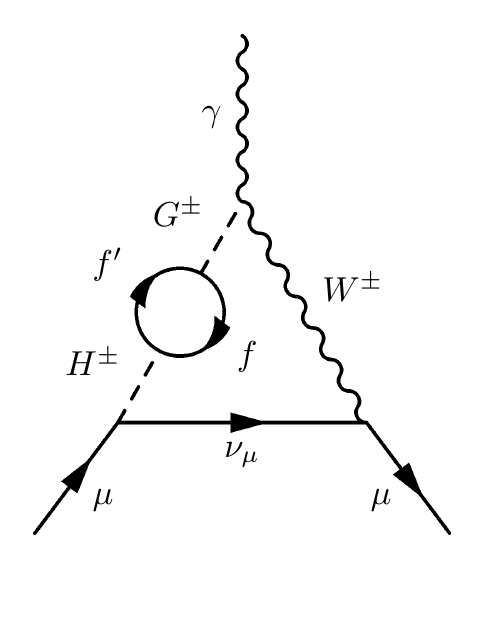}
	\caption{Two-loop fermionic contribution included within \GMTCalc, where $\phi=h,H,A$, $f = u,d,l$, and $f'$ is the $SU(2)_L$ partner of $f$.  We can replace the photon in the left diagram with a $Z$ boson, although these contributions are suppressed.  
		\label{fig:TwoLoopFermionDiagrams}}
\end{figure}

\begin{figure}[tb]
	\centering
	\includegraphics[width=0.32\textwidth]{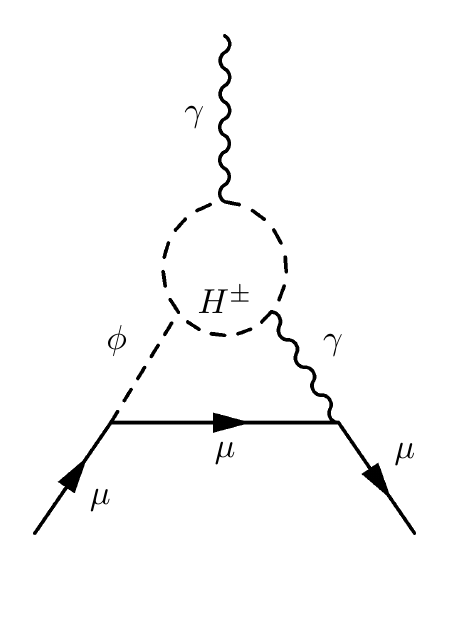}\hfill
	\includegraphics[width=0.32\textwidth]{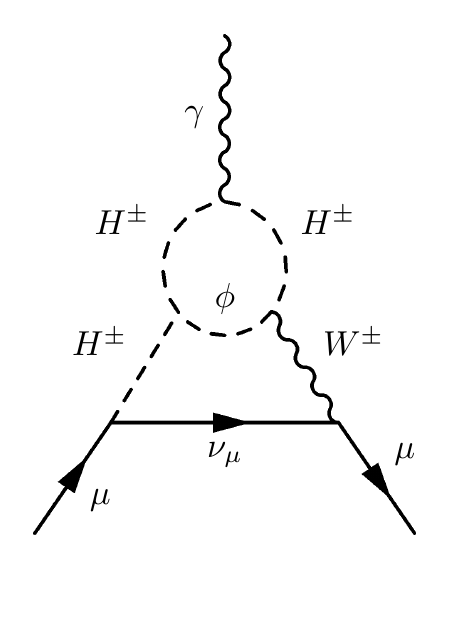}\hfill
	\includegraphics[width=0.32\textwidth]{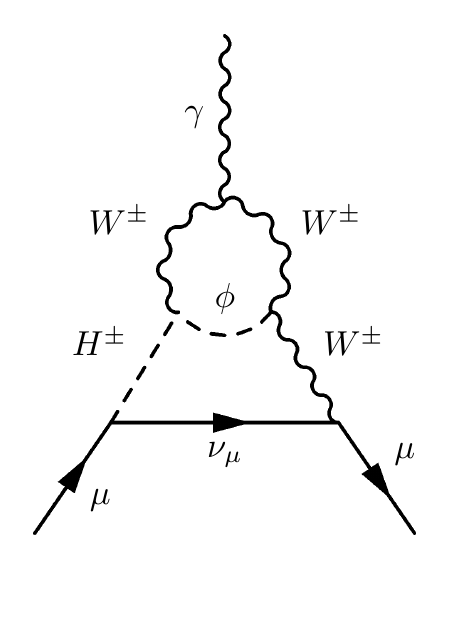}
	\caption{Two-loop bosonic Barr-Zee contribution included within \GMTCalc, where $\phi=h,H,A$.  We can replace the photon in the left diagram with a $Z$ boson, although these contributions are suppressed.  
		\label{fig:TwoLoopBosonicBZDiagrams}}
\end{figure}

\begin{figure}[tb]
	\centering
	\includegraphics[width=0.32\textwidth]{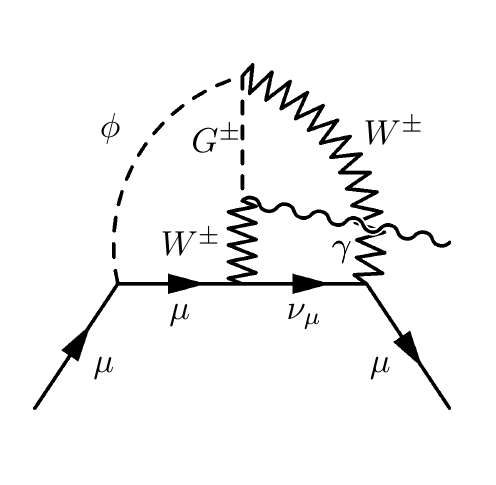}\hfill
	\includegraphics[width=0.32\textwidth]{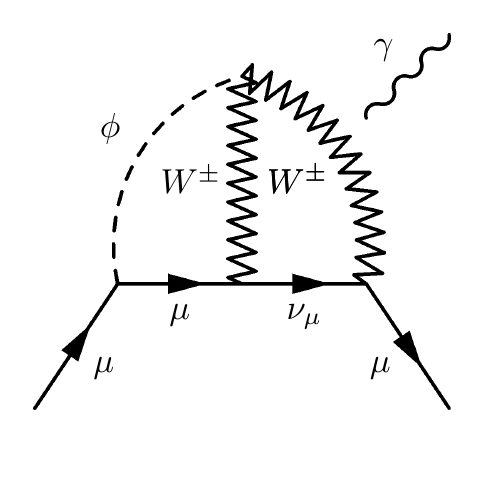}\hfill
	\includegraphics[width=0.32\textwidth]{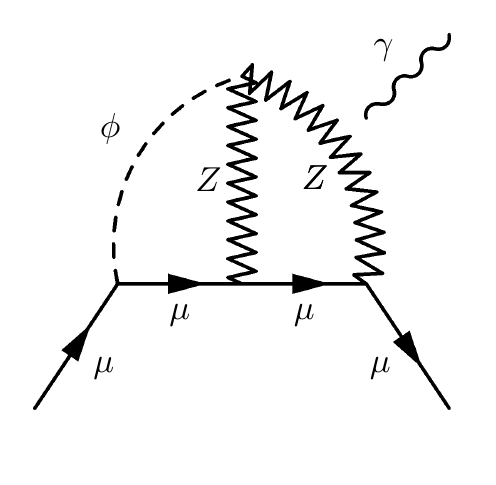}
	\caption{Two-loop bosonic three boson contribution included within \GMTCalc, where $\phi=h,H,A$.  In the middle and right diagrams the photon can couple to either of the vector bosons.  Also in the right diagram the neutral Higgs can be swap its position with one of the $Z$ bosons.  
		\label{fig:TwoLoopBosonic3BDiagrams}}
\end{figure}

The \THDM's state-of-the-art two-loop contributions in \GMTCalc\ are the taken from \cite{Cherchiglia:2016eui}.  The two-loop contributions are divided into fermionic and bosonic parts
\begin{equation} \label{eqn:amutwoloop}
	\amu^{2\ell} = \amuF + \amuB.
\end{equation}
The fermionic contributions are shown in Fig.\ \ref{fig:TwoLoopFermionDiagrams}, and defined in Eq.\ (50) of \cite{Athron:2021evk}.  In \GMTCalc2 we allow for CKM mixing, so in the middle and right Barr-Zee diagrams in the case of $f=u,d$ the fermion loops may have two $SU(2)_L$ partners from different generations.  Again, just as for the one-loop contributions we must remember to subtract off contributions from the \SM-like Higgs to avoid double counting.  The bosonic contributions shown in Figs.\ \ref{fig:TwoLoopBosonicBZDiagrams} and \ref{fig:TwoLoopBosonic3BDiagrams} are defined in Eq.\ (66) of \cite{Athron:2021evk}, and can be divided into Barr-Zee and three boson diagrams respectively.  These contributions have been extended compared to their original mention in \cite{Cherchiglia:2016eui} to allow for a non-zero $\lambda_{6,7}$ in the general scalar potential.  

\GMTCalc\ is available for download from a git [\url{https://github.com/GM2Calc/GM2Calc}] and a hepforge [\url{https://gm2calc.hepforge.org}] repository.  Its requirements have been kept to a minimum, only requiring a \lstinline|C++14| or a \lstinline|C11| compiler as well as the Boost \cite{eigenweb} (version 1.37.0 from [\url{http://eigen.tuxfamily.org}]) and Eigen (version 3.1 from [\url{http://www.boost.org}]) libraries.  To build \GMTCalc\ enter the below commands in the directory where \GMTCalc\ was downloaded
\begin{lstlisting}[language=bash]
	mkdir build; cd build
	cmake ..
	make
\end{lstlisting}

Running \GMTCalc\ can be done straight from the command-line.  Inputs are given through an SLHA-like input file.  Using the provided example file, one can run the command
\begin{lstlisting}[language=bash]
	bin/gm2calc.x --thdm-input-file=../input/example.thdm
\end{lstlisting}
or equivalently replace \lstinline|example.thdm| with the users own input file.  
Alternatively one can use one of \GMTCalc's interfaces.  The C, C++, and Mathematica \cite{Mathematica} interfaces which worked for the \MSSM\ work just as well for the \THDM.  New in \GMTCalc\ version 2 are the \python\ 2 or \python\ 3 \cite{Python} interfaces, which requires the package \lstinline|cppyy| \cite{cppyy} from [\url{https://pypi.org/project/cppyy/}], and attaching the flag \lstinline|-DBUILD_SHARED_LIBS=ON| to the \lstinline|cmake| build step.  


\section{Examples} \label{sec:examples}

In this section we examine some examples of using \GMTCalc2 to calculation the contributions to $\amuBSM$ in the \THDM.  All of these examples are present in Ref.\ \cite{Athron:2021evk}, however we use different interfaces compare to those in the original work.  Each of the examples also uses a scenario taken from the literature, allowing comparison of \GMTCalc2's capabilities with existing tools and techniques. 

\begin{figure}[tb]
	\centering
	\includegraphics[width=0.49\textwidth]{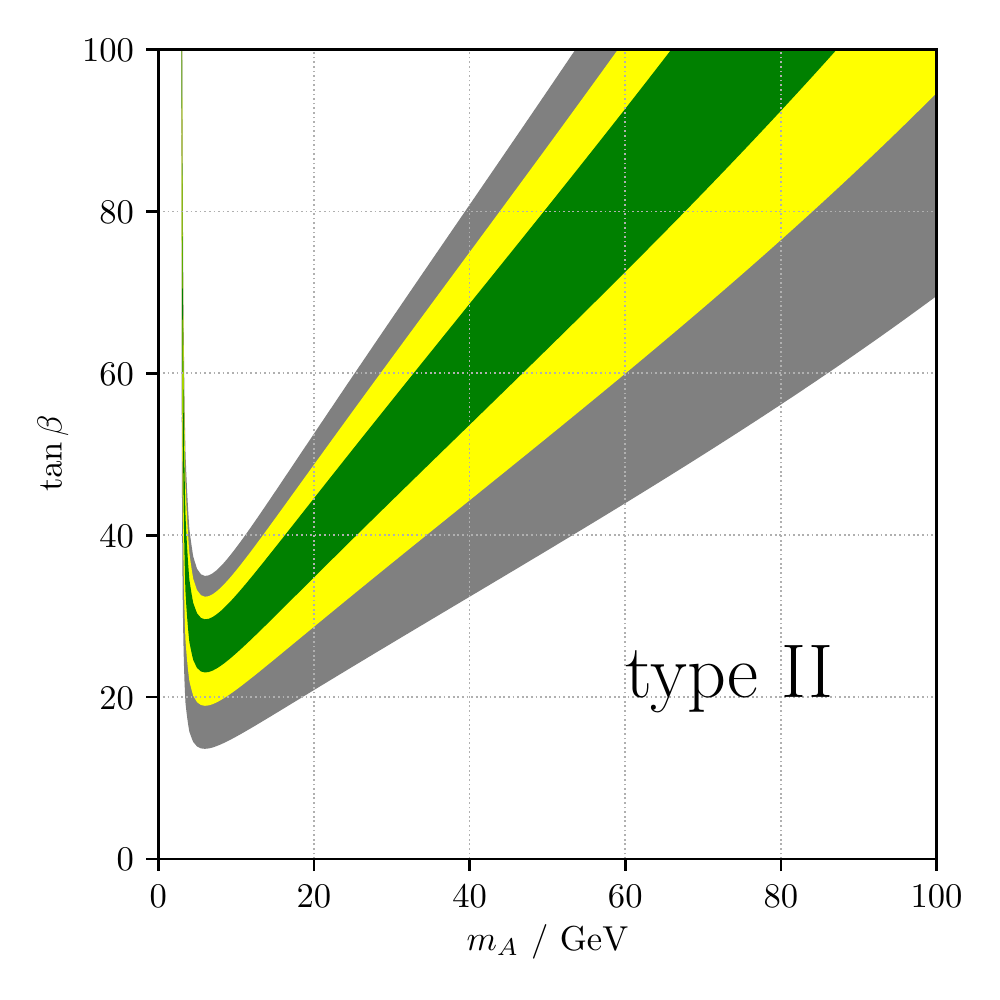}\hfill
	\includegraphics[width=0.49\textwidth]{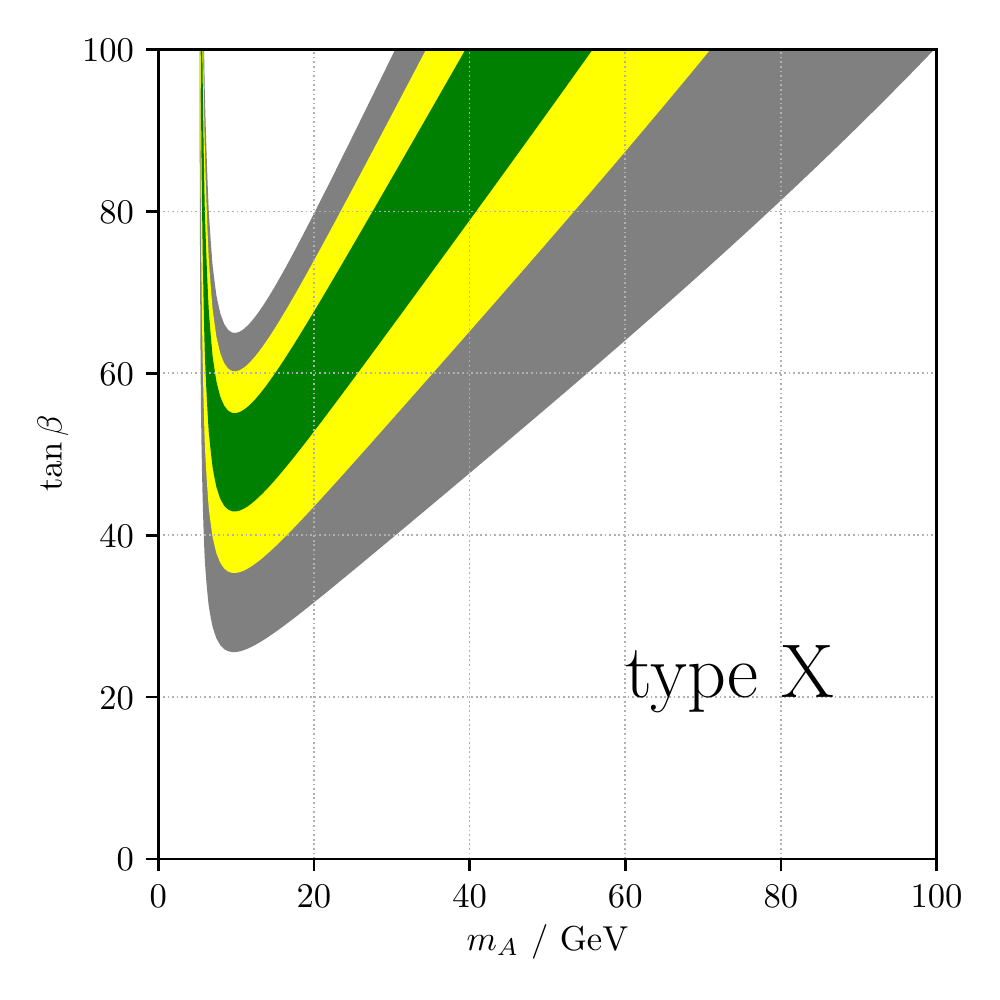}
	\caption{Prediction of the two-loop contributions to $\amuBSM$ from the types-II and X \THDM\, shown in the left plots and right plot respectively.   The contributions are a function of $\tan{(\beta)}$ and $m_A$ with $m_h=126\GeV$, $m_H=m_{H^\pm}=200\GeV$, $\sin(\beta-\alpha)=1$, $\lambda_6=\lambda_7=0$, $m_{12}^2=m_H^2/\tan{(\beta)} + (m_h^2 - \lambda_1 v^2)/\tan{(\beta)}^3$ (see Eq.\ (14) in \cite{Broggio:2014mna}) and $\lambda_1=\sqrt{4\pi}$.  The green, yellow, and grey regions show point which can explain the $\amu$ anomaly in Eq.\ (\ref{eq:amuBSM}) within $1$, $2$, and $3\sigma$ respectively.  
	\label{fig:mA-tb}}
\end{figure}

The first example follows from Ref.\ \cite{Broggio:2014mna} and is a scan over $\tan{\beta}$ and $m_A$ in the types-II and X \THDM.  Below we show the code of the \python\ interface used for the example
\begin{lstlisting}[language=python]
#!/usr/bin/env python

from  __future__ import print_function

from gm2_python_interface import *

cppyy.include(os.path.join("gm2calc","gm2_1loop.hpp"))
cppyy.include(os.path.join("gm2calc","gm2_2loop.hpp"))
cppyy.include(os.path.join("gm2calc","gm2_uncertainty.hpp"))
cppyy.include(os.path.join("gm2calc","gm2_error.hpp"))
cppyy.include(os.path.join("gm2calc","THDM.hpp"))

cppyy.load_library("libgm2calc")

# Load data types
from cppyy.gbl import std
from cppyy.gbl import Eigen
from cppyy.gbl import gm2calc
from cppyy.gbl.gm2calc import SM
from cppyy.gbl.gm2calc import THDM
from cppyy.gbl.gm2calc import Error

# Based on arxiv:1409.3199, examining the mA-tan(beta) plane
def calc_amu(mA, tb, yukawa_type):
	sm = gm2calc.SM()
	sm.set_alpha_em_mz(1.0/128.94579)
	sm.set_mu(2,173.34)
	sm.set_mu(1,1.28)
	sm.set_md(2,4.18)
	sm.set_ml(2,1.77684)
	
	vev = sm.get_v()
	lambda_max = 3.5449077018110321 # Sqrt[4 Pi[]]
	mh = 126.
	mH = 200.
	
	basis = gm2calc.thdm.Mass_basis()
	basis.yukawa_type = gm2calc.thdm.Yukawa_type.type_2
	basis.mh = mh
	basis.mH = mH
	basis.mA = mA
	basis.mHp = mH
	basis.sin_beta_minus_alpha = 1.
	basis.lambda_6 = 0.
	basis.lambda_7 = 0.
	basis.tan_beta = tb
	basis.m122 = mH*mH/tb + (mh*mh - lambda_max*sm.get_v()*sm.get_v())/(tb*tb*tb);
	basis.zeta_u = 0.
	basis.zeta_d = 0.
	basis.zeta_l = 0.
	basis.Delta_u = Eigen.Matrix3d().setZero()
	basis.Delta_d = Eigen.Matrix3d().setZero()
	basis.Delta_l = Eigen.Matrix3d().setZero()
	basis.Pi_u = Eigen.Matrix3d().setZero()
	basis.Pi_d = Eigen.Matrix3d().setZero()
	basis.Pi_l = Eigen.Matrix3d().setZero()
	
	config = gm2calc.thdm.Config()
	config.force_output = False
	config.running_couplings = True
	
	amu = None
	
	try:
		model = gm2calc.THDM(basis,sm,config)
		amu = gm2calc.calculate_amu_1loop(model) + gm2calc.calculate_amu_2loop(model)
	except gm2calc.Error as e:
		pass
	
	return amu
	

mA_start = 1
tb_start = 1
mA_stop = 100
tb_stop = 100
N_steps = 198

print("# mA/GeV            ","# tan(beta)          ","# amu(II)            ","# amu(X)             ")

for i in range(0,N_steps+1):
	for j in range(0,N_steps+1):
		tb = tb_start + i*(tb_stop - tb_start) / N_steps
		mA = mA_start + j*(mA_stop - mA_start) / N_steps
		type_2 = calc_amu(mA, tb, gm2calc.thdm.Yukawa_type.type_2)
		type_X = calc_amu(mA, tb, gm2calc.thdm.Yukawa_type.type_X)

		print("{0:>20.8e} {1:>20.8e} {2:>20.8e} {3:>20.8e}".format(tb,mA,type_2,type_X))
		

\end{lstlisting}
The first 5 lines setup the above \python\ script so that it may be used by both \python2 and \python3, and imports the interface module \lstinline|gm2_python_interface|, which is a convenient script built with \GMTCalc\ which loads the necessary libraries for \GMTCalc\ to work in \python.  Lines 7-11 include the header files needed to run the \THDM\ in \python, while line 13 loads the source code of \GMTCalc.  Next, lines 16-21 load the needed namespaces and classes.  After that is the function \lstinline|calc_amu| which calculates and returns the value of $\amuBSM$, given the inputs $m_A$, $\tan{(\beta)}$ and the \THDM\ type.  Lines 25-30 fix the masses of the \SM\ fermions and the electroweak constant.  Next, lines 32-35 set the needed values to correctly calculate $m_{12}^2$.  Lines 37-56 set up the \lstinline|basis| object, and in this particular example we use the mass basis.  The parameters are set to $m_H=m_{H^\pm}=200$ GeV, $\beta-\alpha=\pi/2$ and are taken from Fig.\ 3 in Ref.\ \cite{Broggio:2014mna}, and on line 47 the value of $m_{12}^2$ is set according to Eq.\ (14) of Ref.\ \cite{Broggio:2014mna}.  Since in this example we will be using the types-II and X \THDM\, we do not need to input values for $\zeta_f$, $\Delta_f$, and $\Pi_f$, as they are ignored, however we show how to input to them for completeness.   Next lines 58-60 set the configuration options, where we choose to use running couplings for the fermion masses.  Finally lines 64-68 calculate the contributions to $\amuBSM$ given the above inputs.  Next in lines 73-77 the parameters of the scan are set.  Then in lines 81-86 the script loops over the values of $m_A$ and $\tan{\beta}$, calculating the contributions to $\amuBSM$ using the function \lstinline|calc_amu|, and finally in line 88 it prints the contribution to $\amuBSM$ in the types-II and X \THDM.  The results of the scan in the above code are shown in Fig.\ \ref{fig:mA-tb}.  

\begin{figure}
	\centering
	\includegraphics[width=0.49\textwidth]{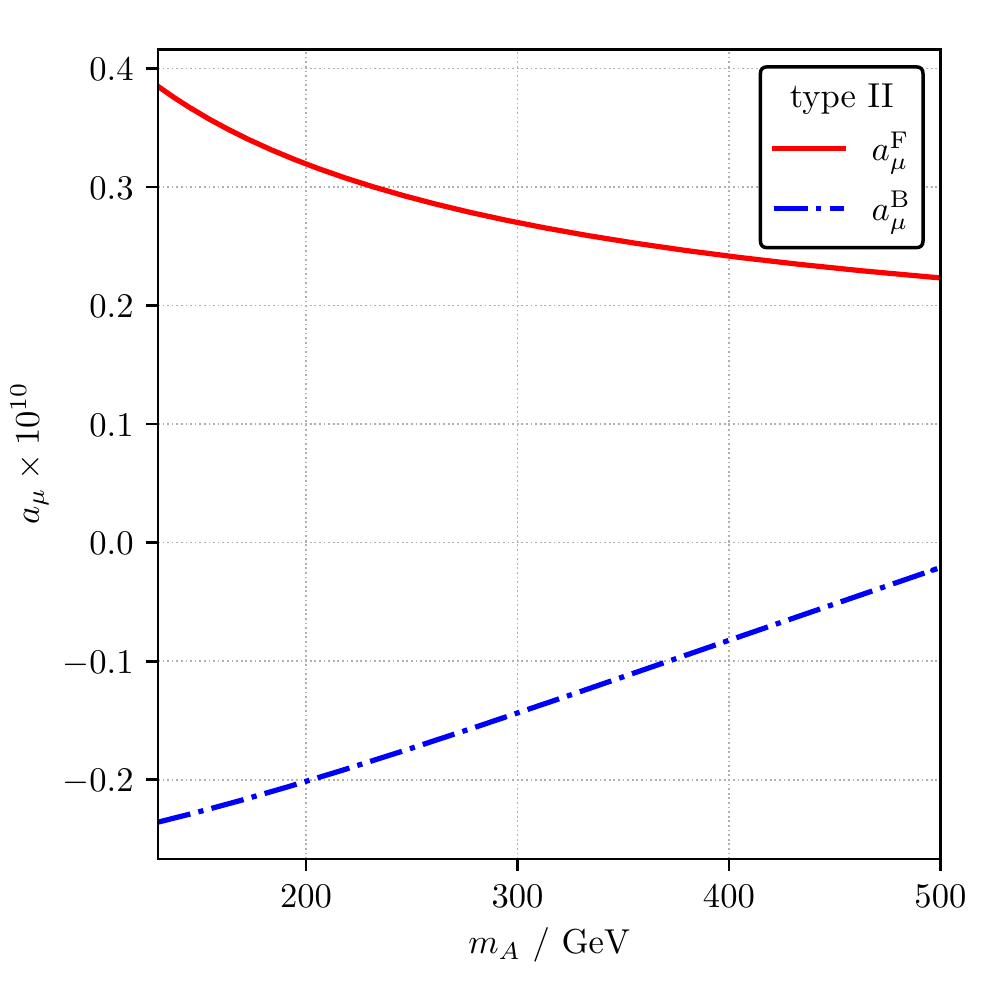}\hfill
	\includegraphics[width=0.49\textwidth]{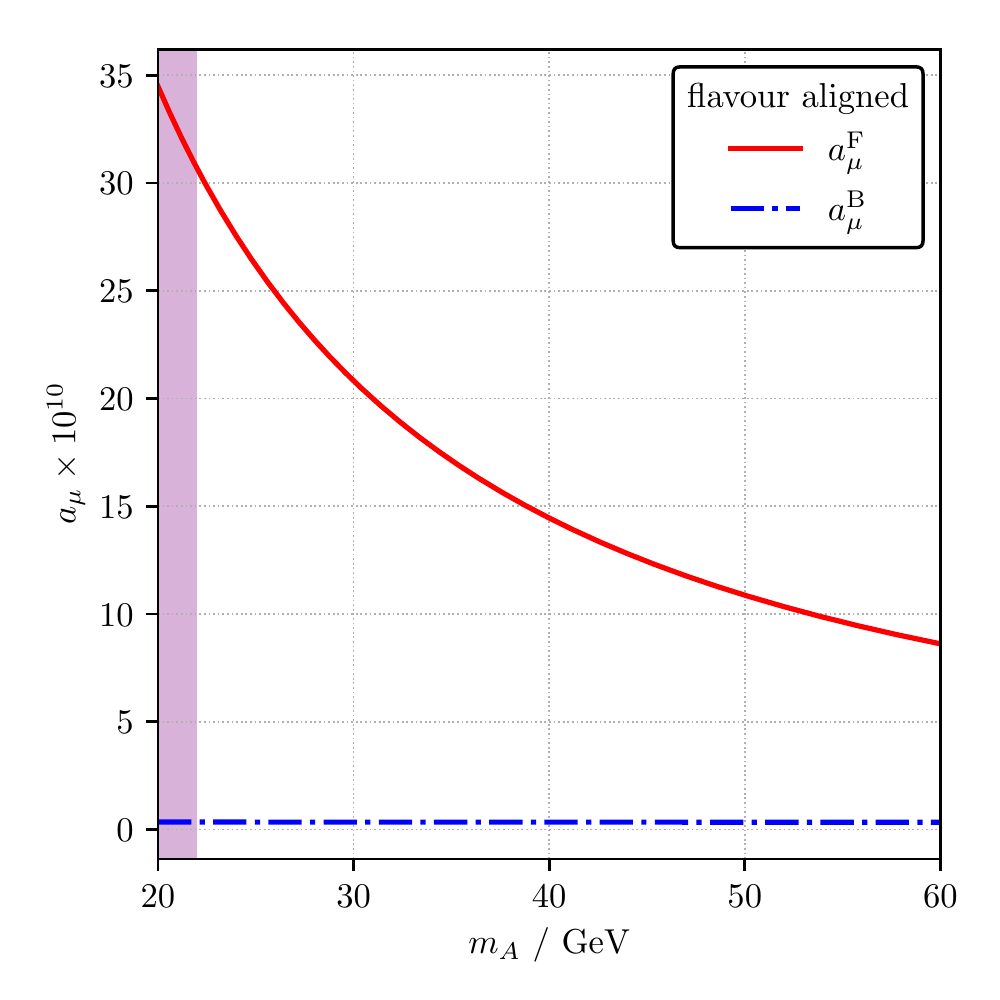}
	\caption{Comparison of the bosonic and fermionic two-loop contributions to $\amuBSM$ in the \THDM, which correspond to the red and blue dashed lines respectively.  The left panel shows a scenario from the type-II \THDM\ from Ref.\ \cite{Eriksson:2009ws}, with the parameters $m_H = 400$ GeV, $m_{H^\pm} = 440$ GeV, $\tan{(\beta)} = 3$, and $m_{12}^2 = 40000$ GeV$^2$.  The right panel shows a scenario in the \FATHDM\ from \figurename~10 in Ref.\ \cite{Cherchiglia:2017uwv}, which uses the parameters $m_H=m_{H^\pm}=150$ GeV, $\zeta_u=\zeta_d=-0.1$, $\zeta_l=50$, $\tan{(\beta)}=2$ and $\sin{(\beta-\alpha)}=0.999$, and fixes $m_{12}^2$ to avoid $h\to A A$ decays.  Also shown in the right panel is a shaded purple region which indicates when the combine one-loop, two-loop fermionic and two-loop bosonic contributions can explain the value of the $\amu$ anomaly in Eq.\ (\ref{eq:amuBSM}).  }
	\label{fig:FB}
\end{figure}

In the next example we compare the size of the bosonic and fermionic two-loop contributions in the type-II and \FATHDM.  This example can be run in \mathematica\ using \texttt{Mathlink}, and the below source code shows the type-II part of the example
\begin{lstlisting}[language=mathematica]
Install["bin/gm2calc.mx"];

CalcAmu2[mA_] :=
{amu2LF, amu2LB} /. GM2CalcAmuTHDMMassBasis[
yukawaType        -> 2,
Mhh               -> { 125, 400 },
MAh               -> mA,
MHp               -> 440,
sinBetaMinusAlpha -> 0.999,
lambda6           -> 0,
lambda7           -> 0,
TB                -> 3,
m122              -> 200^2
];

(* mA values in [130,500] GeV *)
mAValues = Subdivide[130, 500, 200];

(* calculation w/ running couplings *)
GM2CalcSetFlags[runningCouplings -> True];
type2values = CalcAmu2 /@ mAValues;

data2 = { mAValues, type2values[[All,1]], type2values[[All,2]] };

Export["type2.txt", N @ Transpose @ data2, "Table"];
\end{lstlisting}
In the above code line 1 loads the \texttt{MathLink} executable which is built by \GMTCalc\ in its build step.  Lines 3-14 constructs the function \lstinline|CalcAmu2| which takes $m_A$ as input, and returns the individual one-loop, two-loop fermionic and two-loop bosonic contributions.  The parameter choices of $m_H = 400$ GeV, $m_{H^\pm} = 440$ GeV, $\tan{(\beta)} = 3$, $\sin{(\beta-\alpha)}=0.999$, and $m_{12}^2 = 40000$ GeV$^2$ are taken from Ref.\ \cite{Eriksson:2009ws}.  The range of $m_A$ is specified on line 17.  We choose to use running masses for the \SM\ fermions, and then calculate the contributions to $\amuBSM$ on lines 20-21.  Finally in lines 23-25 the data is assembled and outputted.  The results of this scan are shown in the left plot of Fig.\ \ref{fig:FB}.  

The source code for the flavour-aligned part of the example is shown below
\begin{lstlisting}[language=mathematica]
Install["bin/gm2calc.mx"];

CalcAmuFA[mA_] :=
{amu1L, amu2LF, amu2LB} /. GM2CalcAmuTHDMMassBasis[
yukawaType        -> 5,
Mhh               -> { 125, 150 },
MAh               -> mA,
MHp               -> 150,
sinBetaMinusAlpha -> 0.999,
lambda6           -> 0,
lambda7           -> 0,
TB                -> 2,
m122              -> 3187.3 + mA*(3.27803 + 0.0165557*mA),
zetau             -> -0.1,
zetad             -> -0.1,
zetal             -> 50
];

(* mA values in [20,60] GeV *)
mAValues = Subdivide[20, 60, 200];

(* calculation w/ running couplings *)
GM2CalcSetFlags[runningCouplings -> True];
favalues = CalcAmuFA /@ mAValues;

datafa = { mAValues, favalues[[All,1]], favalues[[All,2]], favalues[[All,3]] };

Export["flavouraligned.txt", N @ Transpose @ datafa, "Table"];
\end{lstlisting}
The parameter choices for the above example are taken from Ref.\ \cite{Cherchiglia:2017uwv}, and are chosen to show a region of the \THDM's parameter space where it is possible to explain the value of $\amuBSM$ in Eq.\ (\ref{eq:amuBSM}).  In the function \lstinline|CalcAmuFA| on lines 3-17, we set the parameters to $m_H=m_{H^\pm}=150$ GeV, $\zeta_u=\zeta_d=-0.1$, $\zeta_l=50$, $\tan{(\beta)}=2$, $\sin{(\beta-\alpha)}=0.999$.  We also fix $m_{12}^2=3187.3 + m_A*(3.27803 + 0.0165557*m_A)$ (in GeV$^2$) so there are no signals from $h\to A A$ decays, using Eq.\ (12) from \cite{Cherchiglia:2017uwv}.  The scan for this scenario is over $m_A \in [20,60]$ GeV, and the rest of the script goes similarly to the previous example.  The results of this scan can be seen in the right panel of Fig.\ \ref{fig:FB}.  The shaded purple right region on the left indicate that for low values of $m_A \sim 20$ GeV we can explain the $\amu$ anomaly in the \FATHDM.  

\section{Conclusions} \label{sec:conclusions}

The \THDM\ has been shown to be adept at explaining many of the outstanding anomalies of particle physics, including the $\amu$ anomaly.  \GMTCalc2 provides an easy-to-use and powerful tool which can calculate the contributions to $\amuBSM$ in the types-I, II, X, Y, flavour-aligned, and general $\THDM$.  \GMTCalc2 includes contributions at up to the two-loop level, features CKM mixing, lepton flavour violation, and complex Yukawa couplings, utilizing and building on the calculations done by \cite{Cherchiglia:2016eui}.  We have provided an overview of the code, and given examples of its usage, based on the scenarios from Refs.\ \cite{Broggio:2014mna,Eriksson:2009ws,Cherchiglia:2017uwv}.  Further details can be found in the original work \cite{Athron:2021evk} and on the \GMTCalc\ \href{https://github.com/GM2Calc/GM2Calc}{GitHub page}.  

\section*{Acknowledgments}

We would like to thank the original authors of version 1 of \GMTCalc, including M. Bach, H. G. Fargnoli, C. Gnendiger, R. Greifenhagen, J. Park, and S. Pa{\ss}ehr.  
The research position of D.J. for this work was supported by the Australian Government Research Training Program (RTP) Scholarship and the Deutscher Akademischer Austauschdienst (DAAD) One-Year Research Grant.  The work of P.A. was supported by the Australian Research Council Future Fellowship grant FT160100274.  The work of P.A. and C.B. was also supported with the Australian Research Council Discovery Project grant DP180102209.
This project was undertaken with the assistance of resources and services from the National Computational Infrastructure, which is supported by the Australian Government.  We thank Astronomy Australia Limited for financial support of computing resources.

\bibliographystyle{ieeetr}
\bibliography{paper,TheoryWPbiblio,MuonMoment2HDM,GM2CalcMSSM}

\begin{thebibliography}{10}

\bibitem{Athron:2021evk}
P.~Athron, C.~Balazs, A.~Cherchiglia, D.~H.~J. Jacob, D.~St\"ockinger,
  H.~St\"ockinger-Kim, and A.~Voigt, ``{Two-loop Prediction of the Anomalous
  Magnetic Moment of the Muon in the Two-Higgs Doublet Model with GM2Calc 2},''
  10 2021.

\bibitem{PhysRevLett.126.141801}
B.~Abi {\em et~al.}, ``Measurement of the positive muon anomalous magnetic
  moment to 0.46 ppm,'' {\em Phys. Rev. Lett.}, vol.~126, p.~141801, Apr 2021.

\bibitem{Tanabashi2018}
M.~Tanabashi {\em et~al.}, ``Review of particle physics,'' {\em Phys. Rev. D},
  vol.~98, p.~030001, Aug 2018.

\bibitem{aoyama2020anomalous}
T.~Aoyama {\em et~al.}, ``The anomalous magnetic moment of the muon in the
  standard model,'' {\em Phys. Rept.}, vol.~887, pp.~1--166, 2020.

\bibitem{Czarnecki:2002nt}
A.~Czarnecki, W.~J. Marciano, and A.~Vainshtein, ``{Refinements in electroweak
  contributions to the muon anomalous magnetic moment},'' {\em Phys. Rev.},
  vol.~D67, p.~073006, 2003.
\newblock [Erratum: Phys. Rev. {\bf D73}, 119901 (2006)].

\bibitem{Gnendiger:2013pva}
C.~Gnendiger, D.~St{\"o}ckinger, and H.~St{\"o}ckinger-Kim, ``{The electroweak
  contributions to $(g-2)_\mu$ after the Higgs boson mass measurement},'' {\em
  Phys. Rev.}, vol.~D88, p.~053005, 2013.

\bibitem{Davier:2017zfy}
M.~Davier, A.~Hoecker, B.~Malaescu, and Z.~Zhang, ``{Reevaluation of the
  hadronic vacuum polarisation contributions to the Standard Model predictions
  of the muon $g-2$ and ${\alpha (m_Z^2)}$ using newest hadronic cross-section
  data},'' {\em Eur. Phys. J.}, vol.~C77, no.~12, p.~827, 2017.

\bibitem{Keshavarzi:2018mgv}
A.~Keshavarzi, D.~Nomura, and T.~Teubner, ``{Muon $g-2$ and $\alpha(M_Z^2)$: a
  new data-based analysis},'' {\em Phys. Rev.}, vol.~D97, no.~11, p.~114025,
  2018.

\bibitem{Colangelo:2018mtw}
G.~Colangelo, M.~Hoferichter, and P.~Stoffer, ``{Two-pion contribution to
  hadronic vacuum polarization},'' {\em JHEP}, vol.~02, p.~006, 2019.

\bibitem{Hoferichter:2019gzf}
M.~Hoferichter, B.-L. Hoid, and B.~Kubis, ``{Three-pion contribution to
  hadronic vacuum polarization},'' {\em JHEP}, vol.~08, p.~137, 2019.

\bibitem{Davier:2019can}
M.~Davier, A.~Hoecker, B.~Malaescu, and Z.~Zhang, ``{A new evaluation of the
  hadronic vacuum polarisation contributions to the muon anomalous magnetic
  moment and to $\mathbf{\boldsymbol\alpha(m_Z^2)}$},'' {\em Eur. Phys. J.},
  vol.~C80, no.~3, p.~241, 2020.

\bibitem{Keshavarzi:2019abf}
A.~Keshavarzi, D.~Nomura, and T.~Teubner, ``{The $g-2$ of charged leptons,
  $\alpha(M_Z^2)$ and the hyperfine splitting of muonium},'' {\em Phys. Rev.},
  vol.~D101, p.~014029, 2020.

\bibitem{Kurz:2014wya}
A.~Kurz, T.~Liu, P.~Marquard, and M.~Steinhauser, ``{Hadronic contribution to
  the muon anomalous magnetic moment to next-to-next-to-leading order},'' {\em
  Phys. Lett.}, vol.~B734, pp.~144--147, 2014.

\bibitem{Melnikov:2003xd}
K.~Melnikov and A.~Vainshtein, ``{Hadronic light-by-light scattering
  contribution to the muon anomalous magnetic moment revisited},'' {\em Phys.
  Rev.}, vol.~D70, p.~113006, 2004.

\bibitem{Masjuan:2017tvw}
P.~Masjuan and P.~S{\'a}nchez-Puertas, ``{Pseudoscalar-pole contribution to the
  $(g_{\mu}-2)$: a rational approach},'' {\em Phys. Rev.}, vol.~D95, no.~5,
  p.~054026, 2017.

\bibitem{Colangelo:2017fiz}
G.~Colangelo, M.~Hoferichter, M.~Procura, and P.~Stoffer, ``{Dispersion
  relation for hadronic light-by-light scattering: two-pion contributions},''
  {\em JHEP}, vol.~04, p.~161, 2017.

\bibitem{Hoferichter:2018kwz}
M.~Hoferichter, B.-L. Hoid, B.~Kubis, S.~Leupold, and S.~P. Schneider,
  ``{Dispersion relation for hadronic light-by-light scattering: pion pole},''
  {\em JHEP}, vol.~10, p.~141, 2018.

\bibitem{Gerardin:2019vio}
A.~G{\'e}rardin, H.~B. Meyer, and A.~Nyffeler, ``{Lattice calculation of the
  pion transition form factor with $N_f=2+1$ Wilson quarks},'' {\em Phys.
  Rev.}, vol.~D100, no.~3, p.~034520, 2019.

\bibitem{Bijnens:2019ghy}
J.~Bijnens, N.~Hermansson-Truedsson, and A.~Rodr{\'i}guez-S{\'a}nchez,
  ``{Short-distance constraints for the HLbL contribution to the muon anomalous
  magnetic moment},'' {\em Phys. Lett.}, vol.~B798, p.~134994, 2019.

\bibitem{Colangelo:2019uex}
G.~Colangelo, F.~Hagelstein, M.~Hoferichter, L.~Laub, and P.~Stoffer,
  ``{Longitudinal short-distance constraints for the hadronic light-by-light
  contribution to $(g-2)_\mu$ with large-$N_c$ Regge models},'' {\em JHEP},
  vol.~03, p.~101, 2020.

\bibitem{Pauk:2014rta}
V.~Pauk and M.~Vanderhaeghen, ``{Single meson contributions to the muon`s
  anomalous magnetic moment},'' {\em Eur. Phys. J.}, vol.~C74, no.~8, p.~3008,
  2014.

\bibitem{Danilkin:2016hnh}
I.~Danilkin and M.~Vanderhaeghen, ``{Light-by-light scattering sum rules in
  light of new data},'' {\em Phys. Rev.}, vol.~D95, no.~1, p.~014019, 2017.

\bibitem{Jegerlehner:2017gek}
F.~Jegerlehner, ``{The Anomalous Magnetic Moment of the Muon},'' {\em Springer
  Tracts Mod. Phys.}, vol.~274, pp.~1--693, 2017.

\bibitem{Knecht:2018sci}
M.~Knecht, S.~Narison, A.~Rabemananjara, and D.~Rabetiarivony, ``{Scalar meson
  contributions to $a_\mu$ from hadronic light-by-light scattering},'' {\em
  Phys. Lett.}, vol.~B787, pp.~111--123, 2018.

\bibitem{Eichmann:2019bqf}
G.~Eichmann, C.~S. Fischer, and R.~Williams, ``{Kaon-box contribution to the
  anomalous magnetic moment of the muon},'' {\em Phys. Rev.}, vol.~D101, no.~5,
  p.~054015, 2020.

\bibitem{Roig:2019reh}
P.~Roig and P.~S{\'a}nchez-Puertas, ``{Axial-vector exchange contribution to
  the hadronic light-by-light piece of the muon anomalous magnetic moment},''
  {\em Phys. Rev.}, vol.~D101, no.~7, p.~074019, 2020.

\bibitem{Blum:2019ugy}
T.~Blum, N.~Christ, M.~Hayakawa, T.~Izubuchi, L.~Jin, C.~Jung, and C.~Lehner,
  ``{The hadronic light-by-light scattering contribution to the muon anomalous
  magnetic moment from lattice QCD},'' {\em Phys. Rev. Lett.}, vol.~124,
  no.~13, p.~132002, 2020.

\bibitem{Colangelo:2014qya}
G.~Colangelo, M.~Hoferichter, A.~Nyffeler, M.~Passera, and P.~Stoffer,
  ``{Remarks on higher-order hadronic corrections to the muon $g-2$},'' {\em
  Phys. Lett.}, vol.~B735, pp.~90--91, 2014.

\bibitem{Czarnecki2001}
A.~Czarnecki and W.~J. Marciano, ``Muon anomalous magnetic moment: A harbinger
  for "new physics'','' {\em Phys. Rev. D}, vol.~64, p.~013014, Jun 2001.

\bibitem{Athron:2021iuf}
P.~{Athron}, C.~{Bal{\'a}zs}, D.~H.~J. {Jacob}, W.~{Kotlarski},
  D.~{St{\"o}ckinger}, and H.~{St{\"o}ckinger-Kim}, ``{New physics explanations
  of a$_{{\ensuremath{\mu}}}$ in light of the FNAL muon g {\ensuremath{-}} 2
  measurement},'' {\em Journal of High Energy Physics}, vol.~2021, p.~80, Sept.
  2021.

\bibitem{Martin1997}
S.~P. Martin, ``{A Supersymmetry primer},'' pp.~1--98, 1997.
\newblock [Adv. Ser. Direct. High Energy Phys.18,1(1998)].

\bibitem{Athron:2021auq}
P.~Athron, C.~Balazs, T.~E. Gonzalo, D.~Jacob, F.~Mahmoudi, and C.~Sierra,
  ``{Likelihood analysis of the flavour anomalies and g \textendash{} 2 in the
  general two Higgs doublet model},'' {\em JHEP}, vol.~01, p.~037, 2022.

\bibitem{Haber:2015pua}
H.~E. Haber and O.~St\r{a}l, ``{New LHC benchmarks for the $\mathcal{CP}$
  -conserving two-Higgs-doublet model},'' {\em Eur. Phys. J. C}, vol.~75,
  no.~10, p.~491, 2015.
\newblock [Erratum: Eur.Phys.J.C 76, 312 (2016)].

\bibitem{Baglio:2014nea}
J.~Baglio, O.~Eberhardt, U.~Nierste, and M.~Wiebusch, ``{Benchmarks for Higgs
  Pair Production and Heavy Higgs boson Searches in the Two-Higgs-Doublet Model
  of Type II},'' {\em Phys. Rev. D}, vol.~90, no.~1, p.~015008, 2014.

\bibitem{Eberhardt:2013uba}
O.~Eberhardt, U.~Nierste, and M.~Wiebusch, ``{Status of the two-Higgs-doublet
  model of type II},'' {\em JHEP}, vol.~07, p.~118, 2013.

\bibitem{Chowdhury:2015yja}
D.~Chowdhury and O.~Eberhardt, ``{Global fits of the two-loop renormalized
  Two-Higgs-Doublet model with soft Z$_{2}$ breaking},'' {\em JHEP}, vol.~11,
  p.~052, 2015.

\bibitem{Chowdhury:2017aav}
D.~Chowdhury and O.~Eberhardt, ``{Update of Global Two-Higgs-Doublet Model
  Fits},'' {\em JHEP}, vol.~05, p.~161, 2018.

\bibitem{Misiak:2017bgg}
M.~Misiak and M.~Steinhauser, ``{Weak radiative decays of the B meson and
  bounds on $M_{H^\pm }$ in the Two-Higgs-Doublet Model},'' {\em Eur. Phys. J.
  C}, vol.~77, no.~3, p.~201, 2017.

\bibitem{Hu:2016gpe}
Q.-Y. Hu, X.-Q. Li, and Y.-D. Yang, ``{$B^0\to K^{\ast 0}\mu^+\mu^-$ decay in
  the Aligned Two-Higgs-Doublet Model},'' {\em Eur. Phys. J. C}, vol.~77,
  no.~3, p.~190, 2017.

\bibitem{Li:2014fea}
X.-Q. Li, J.~Lu, and A.~Pich, ``{$B_{s,d}^0 \to \ell^+\ell^-$ Decays in the
  Aligned Two-Higgs-Doublet Model},'' {\em JHEP}, vol.~06, p.~022, 2014.

\bibitem{Enomoto:2015wbn}
T.~Enomoto and R.~Watanabe, ``{Flavor constraints on the Two Higgs Doublet
  Models of Z$_{2}$ symmetric and aligned types},'' {\em JHEP}, vol.~05,
  p.~002, 2016.

\bibitem{Arnan:2017lxi}
P.~Arnan, D.~Be\v{c}irevi\'c, F.~Mescia, and O.~Sumensari, ``{Two Higgs doublet
  models and $b\rightarrow s$ exclusive decays},'' {\em Eur. Phys. J. C},
  vol.~77, no.~11, p.~796, 2017.

\bibitem{Krause:2016oke}
M.~Krause, R.~Lorenz, M.~Muhlleitner, R.~Santos, and H.~Ziesche,
  ``{Gauge-independent Renormalization of the 2-Higgs-Doublet Model},'' {\em
  JHEP}, vol.~09, p.~143, 2016.

\bibitem{Krause:2016xku}
M.~Krause, M.~Muhlleitner, R.~Santos, and H.~Ziesche, ``{Higgs-to-Higgs boson
  decays in a 2HDM at next-to-leading order},'' {\em Phys. Rev. D}, vol.~95,
  no.~7, p.~075019, 2017.

\bibitem{Altenkamp:2017kxk}
L.~Altenkamp, S.~Dittmaier, and H.~Rzehak, ``{Precision calculations for $h \to
  WW/ZZ \to 4$ fermions in the Two-Higgs-Doublet Model with Prophecy4f},'' {\em
  JHEP}, vol.~03, p.~110, 2018.

\bibitem{Altenkamp:2017ldc}
L.~Altenkamp, S.~Dittmaier, and H.~Rzehak, ``{Renormalization schemes for the
  Two-Higgs-Doublet Model and applications to h \textrightarrow{} WW/ZZ
  \textrightarrow{} 4 fermions},'' {\em JHEP}, vol.~09, p.~134, 2017.

\bibitem{Altenkamp:2018hrq}
L.~Altenkamp, M.~Boggia, S.~Dittmaier, and H.~Rzehak, ``{Electroweak
  corrections in the Two-Higgs-Doublet Model and Singlet Extension of the
  Standard Model},'' {\em PoS}, vol.~LL2018, p.~011, 2018.

\bibitem{Denner:2018opp}
A.~Denner, S.~Dittmaier, and J.-N. Lang, ``{Renormalization of mixing
  angles},'' {\em JHEP}, vol.~11, p.~104, 2018.

\bibitem{Kanemura:2018yai}
S.~Kanemura, M.~Kikuchi, K.~Mawatari, K.~Sakurai, and K.~Yagyu, ``{Loop effects
  on the Higgs decay widths in extended Higgs models},'' {\em Phys. Lett. B},
  vol.~783, pp.~140--149, 2018.

\bibitem{Krause:2019qwe}
M.~Krause and M.~M\"uhlleitner, ``{Impact of Electroweak Corrections on Neutral
  Higgs Boson Decays in Extended Higgs Sectors},'' {\em JHEP}, vol.~04, p.~083,
  2020.

\bibitem{Aiko:2021can}
M.~Aiko, S.~Kanemura, and K.~Sakurai, ``{Radiative corrections to decays of
  charged Higgs bosons in two Higgs doublet models},'' 8 2021.

\bibitem{Jueid:2021avn}
A.~Jueid, J.~Kim, S.~Lee, and J.~Song, ``{Type-X two Higgs doublet model in
  light of the muon $\mathbf{g-2}$: confronting Higgs and collider data},'' 4
  2021.

\bibitem{Han:2021gfu}
X.-F. Han, T.~Li, H.-X. Wang, L.~Wang, and Y.~Zhang, ``{Lepton-specific inert
  two-Higgs-doublet model confronted with the new results for muon and electron
  g-2 anomalies and multilepton searches at the LHC},'' {\em Phys. Rev. D},
  vol.~104, no.~11, p.~115001, 2021.

\bibitem{Dey:2021pyn}
A.~Dey, J.~Lahiri, and B.~Mukhopadhyaya, ``{Muon g-2 and a type-X two Higgs
  doublet scenario: some studies in high-scale validity},'' 6 2021.

\bibitem{Hou:2021qmf}
W.-S. Hou and G.~Kumar, ``{Charged lepton flavor violation in light of Muon
  $g-2$},'' 7 2021.

\bibitem{Hou:2021wjj}
W.-S. Hou, ``{Decadal Mission for the New Physics Higgs/Flavor Era},'' 9 2021.

\bibitem{Han:2016gvr}
C.~Han, K.-i. Hikasa, L.~Wu, J.~M. Yang, and Y.~Zhang, ``{Status of CMSSM in
  light of current LHC Run-2 and LUX data},'' {\em Phys. Lett. B}, vol.~769,
  pp.~470--476, 2017.

\bibitem{Ren:2017ymm}
J.~Ren, L.~Wu, J.~M. Yang, and J.~Zhao, ``{Exploring supersymmetry with machine
  learning},'' {\em Nucl. Phys. B}, vol.~943, p.~114613, 2019.

\bibitem{Endo:2020mqz}
M.~Endo, K.~Hamaguchi, S.~Iwamoto, and T.~Kitahara, ``{Muon $g$\textendash{}$2$
  vs LHC Run 2 in supersymmetric models},'' {\em JHEP}, vol.~04, p.~165, 2020.

\bibitem{Chakraborti:2020vjp}
M.~Chakraborti, S.~Heinemeyer, and I.~Saha, ``{Improved $(g-2)_\mu$
  Measurements and Supersymmetry},'' {\em Eur. Phys. J. C}, vol.~80, no.~10,
  p.~984, 2020.

\bibitem{Chakraborti:2021kkr}
M.~Chakraborti, S.~Heinemeyer, and I.~Saha, ``{Improved ${(g-2)_\mu }$
  measurements and wino/higgsino dark matter},'' {\em Eur. Phys. J. C},
  vol.~81, no.~12, p.~1069, 2021.

\bibitem{Chakraborti:2021dli}
M.~Chakraborti, S.~Heinemeyer, and I.~Saha, ``{The new \textquotedblleft{}MUON
  G-2\textquotedblright{} result and supersymmetry},'' {\em Eur. Phys. J. C},
  vol.~81, no.~12, p.~1114, 2021.

\bibitem{Wang:2021bcx}
F.~Wang, L.~Wu, Y.~Xiao, J.~M. Yang, and Y.~Zhang, ``{GUT-scale constrained
  SUSY in light of new muon g-2 measurement},'' {\em Nucl. Phys. B}, vol.~970,
  p.~115486, 2021.

\bibitem{Endo:2021zal}
M.~Endo, K.~Hamaguchi, S.~Iwamoto, and T.~Kitahara, ``{Supersymmetric
  interpretation of the muon g \textendash{} 2 anomaly},'' {\em JHEP}, vol.~07,
  p.~075, 2021.

\bibitem{Ibe:2021cvf}
M.~Ibe, S.~Kobayashi, Y.~Nakayama, and S.~Shirai, ``{Muon $g-2$ in Gauge
  Mediation without SUSY CP Problem},'' 4 2021.

\bibitem{VanBeekveld:2021tgn}
M.~Van~Beekveld, W.~Beenakker, M.~Schutten, and J.~De~Wit, ``{Dark matter,
  fine-tuning and $(g-2)_{\mu}$ in the pMSSM},'' {\em SciPost Phys.}, vol.~11,
  no.~3, p.~049, 2021.

\bibitem{Aboubrahim:2021xfi}
A.~Aboubrahim, M.~Klasen, and P.~Nath, ``{What the Fermilab muon $g-$2
  experiment tells us about discovering supersymmetry at high luminosity and
  high energy upgrades to the LHC},'' {\em Phys. Rev. D}, vol.~104, no.~3,
  p.~035039, 2021.

\bibitem{Chakraborti:2021bmv}
M.~Chakraborti, L.~Roszkowski, and S.~Trojanowski, ``{GUT-constrained
  supersymmetry and dark matter in light of the new $(g-2)_\mu$
  determination},'' {\em JHEP}, vol.~05, p.~252, 2021.

\bibitem{Jeong:2021qey}
K.~S. Jeong, J.~Kawamura, and C.~B. Park, ``{Mixed modulus and anomaly
  mediation in light of the muon g \ensuremath{-} 2 anomaly},'' {\em JHEP},
  vol.~10, p.~064, 2021.

\bibitem{Lamborn:2021snt}
J.~L. Lamborn, T.~Li, J.~A. Maxin, and D.~V. Nanopoulos, ``{Resolving the (g
  \ensuremath{-} 2)$_{\mu}$ discrepancy with $ \mathcal{F} $\textendash{}SU(5)
  intersecting D-branes},'' {\em JHEP}, vol.~11, p.~081, 2021.

\bibitem{Cornell:2016gho}
J.~M. Cornell, ``{Global fits of scalar singlet dark matter with GAMBIT},''
  {\em PoS}, vol.~ICHEP2016, p.~118, 2016.

\bibitem{GAMBIT:2017zdo}
P.~Athron {\em et~al.}, ``{A global fit of the MSSM with GAMBIT},'' {\em Eur.
  Phys. J. C}, vol.~77, no.~12, p.~879, 2017.

\bibitem{GAMBIT:2017snp}
P.~Athron {\em et~al.}, ``{Global fits of GUT-scale SUSY models with GAMBIT},''
  {\em Eur. Phys. J. C}, vol.~77, no.~12, p.~824, 2017.

\bibitem{Abdughani:2019wuv}
M.~Abdughani, J.~Ren, L.~Wu, J.~M. Yang, and J.~Zhao, ``{Supervised deep
  learning in high energy phenomenology: a mini review},'' {\em Commun. Theor.
  Phys.}, vol.~71, no.~8, p.~955, 2019.

\bibitem{Cherchiglia:2016eui}
A.~Cherchiglia, P.~Kneschke, D.~St\"ockinger, and H.~St\"ockinger-Kim, ``{The
  muon magnetic moment in the 2HDM: complete two-loop result},'' {\em JHEP},
  vol.~01, p.~007, 2017.

\bibitem{eigenweb}
G.~Guennebaud, B.~Jacob, {\em et~al.}, ``Eigen v3.''
  \url{http://eigen.tuxfamily.org}, 2010.

\bibitem{Mathematica}
W.~R. Inc., ``Mathematica.''
\newblock Champaign, IL.

\bibitem{Python}
G.~van Rossum, ``Python.'' \url{http://www.python.org}.

\bibitem{cppyy}
W.~T. Lavrijsen and A.~Dutta, ``{High-performance Python--C++ bindings with
  PyPy and Cling}.''
  \url{http://cern.ch/wlav/Cppyy_LavrijsenDutta_PyHPC16.pdf}, 2018.

\bibitem{Broggio:2014mna}
A.~Broggio, E.~J. Chun, M.~Passera, K.~M. Patel, and S.~K. Vempati, ``{Limiting
  two-Higgs-doublet models},'' {\em JHEP}, vol.~11, p.~058, 2014.

\bibitem{Eriksson:2009ws}
D.~Eriksson, J.~Rathsman, and O.~Stal, ``{2HDMC: Two-Higgs-Doublet Model
  Calculator Physics and Manual},'' {\em Comput. Phys. Commun.}, vol.~181,
  pp.~189--205, 2010.

\bibitem{Cherchiglia:2017uwv}
A.~Cherchiglia, D.~Stöckinger, and H.~Stöckinger-Kim, ``{Muon g-2 in the
  2HDM: maximum results and detailed phenomenology},'' {\em Phys. Rev. D},
  vol.~98, p.~035001, 2018.

\end{thebibliography}

\end{document}